\documentstyle[sprocl]{article}

\input{psfig}
\def\NPB{{\em Nucl. Phys.} B}
\def\PLB{{\em Phys. Lett.}  B}
\def\PRL{\em Phys. Rev. Lett.}
\def\PRD{{\em Phys. Rev.} D}

\def\be{\begin{equation}}
\def\nn{\noindent}
\def\ie{{\it i.e.}}

\def\etal{{\it et al.}}
\def\ee{\end{equation}}
\def\bea{\begin{eqnarray}}
\def\eea{\end{eqnarray}}

\begin{document}

\rightline{\vbox{\halign{&#\hfil\cr
SLAC-PUB-8194\cr
July 1999\cr}}}
\vspace{0.8in}

\title{{DISTINGUISHING INDIRECT SIGNATURES ARISING FROM NEW PHYSICS AT THE NLC}
\footnote{To appear in the {\it Proceedings of the World-Wide Study of 
Physics and Detectors for Future Linear Colliders(LCWS99)}, Sitges, 
Barcelona, Spain, 28 April-5 May 1999}
}

\author{ {T.G. RIZZO}
\footnote{Work supported by the Department of Energy, 
Contract DE-AC03-76SF00515}
}

\address{Stanford Linear Accelerator Center,\\
Stanford University, Stanford, CA 94309, USA}


\maketitle\abstracts{Many sources of new physics can lead to shifts in the 
Standard Model predictions for cross sections and asymmetries at the NLC below 
their direct production thresholds. In this talk we discuss some of the tools 
that are useful for distinguishing amongst these new physics scenarios. 
$R$-parity violation and extensions of the Standard Model gauge structure 
are two typical non-minimal realizations of supersymmetry which provide us 
with an important test case to examine.}

While the MSSM provides a simplified framework in which to work, most 
would agree that the MSSM is itself inadequate due to the very large 
number of free parameters it contains. 
In going beyond the MSSM there are many possible paths to follow. In this 
talk we discuss two of the simplest of these scenarios: an extension of the 
SM gauge group by an additional $U(1)$ factor broken near the TeV scale and 
$R$-parity violation, both of which are well-motivated by string theory.  
Although these two alternatives would appear to have little in 
common they can lead to similar phenomenology at 
future linear colliders and may be easily confused in certain regions of 
the parameter space for each class of model. This is a particular example of a 
more general situation wherein various distinct classes of new physics models 
can lead to similar experimental signatures at colliders 
that differ only in detail. The purpose of the present 
analysis is to explore the tools that can be used to distinguish these 
scenarios at $e^+e^-$ colliders~{\cite {tgr}} which can be then applied to 
other more complex scenarios~{\cite {jlh}}.

If $R$-parity is violated it is 
possible that the exchange of sparticles can contribute significantly to 
SM processes and may even be produced as bumps~{\cite {bumps}} in cross 
sections if they are kinematically 
accessible. Below threshold, these new spin-0 exchanges may make 
their presence known via indirect effects on cross sections and 
other observables even when they occur in the 
$t$- or $u$-channels. Here we will address the question of whether the 
effects of the exchange of such particles can be differentiated from those 
conventionally associated with a $Z'$ below threshold at linear colliders. 
If just the $R$-parity violating 
$\lambda LLE^c$ terms of the superpotential are present it is clear that only 
the observables associated with 
leptonic processes will be affected by the exchange of $\tilde \nu$'s in the 
($i$) $t$- or ($ii$) $s$-channels (in the case of 
$e^+e^- \to \mu^+\mu^-/ \tau^+\tau$) or ($iii$) both (for the case of 
Bhabha scattering). [The generalization to include hadronic final states is 
reasonably straightforward.] 
In making the $Z'$ vs $\tilde \nu$ comparison it is 
important to remember that in the $Z'$ case, for the $\mu^+\mu^-$ and 
$ \tau^+\tau$ final states, the angular distributions retain 
their SM forms, $\sim A(1+z)^2+B(1-z)^2$, where $z=\cos \theta$, but with 
the $s$-dependent constants $A,B$ get  shifted from their SM values. In the 
case $\tilde \nu$ exchange the angular distribution is modified in a more 
complex manner. It is also important to remember that we must in principle \
allow the $Z'$ couplings to fermions, $g_{L,R}'^f$, to be completely arbitrary 
in our discussion below in order to avoid any model dependence. We note that 
in all cases the single beam polarization asymmetry, $A_{LR}$, does not help 
to separate these two new physics sources when only leptonic final states 
are involved.

In case ($i$) the $\tilde \nu$ exchange leaves the SM 
value of $A$ unchanged while adding a $z$-dependent contribution to $B$. By 
contrast a generic $Z'$ will modify both $A$ and $B$ leading to distortions 
in the angular distribution in both the forward and backward directions 
depending upon the details of the $Z'$ couplings. 
Fig. 1 shows the resulting change in the distribution for 
$\tilde \nu$ exchange; note that the 
cross section in the forward direction is left unaltered. Fig. 2 shows the 
corresponding $Z'$ induced shifts for comparison. 
Fig. 1 also shows 
that fits to $A,B$ should be able to isolate the $\tilde \nu$ scenario 
except in conspiratorial parameter space regions with suitably chosen $Z'$ 
couplings. With luminosities of 150 $fb^{-1}$, except for these conspiratorial 
cases, the two scenarios remain separable up to $m_{\tilde \nu} \simeq 
(7-8)\lambda$ TeV. We note that with only low statistics 
the $z$-dependence of $B$ may not be visible and only a general average shift 
in its value is obtained from the fit. At some point however, enough statistics 
can be accumulated such that fits with $B$ constant give bad $\chi^2$'s. For 
a luminosity of 200 $fb^{-1}$ at a TeV collider we estimate that this occurs 
when $m_{\tilde \nu} \leq 2.4\lambda$ TeV. Though the reach is much smaller, 
in this case there is no confusion with the $Z'$ scenario.

\vspace*{-0.5cm}
\nn
\begin{figure}[htbp]
\centerline{
\psfig{figure=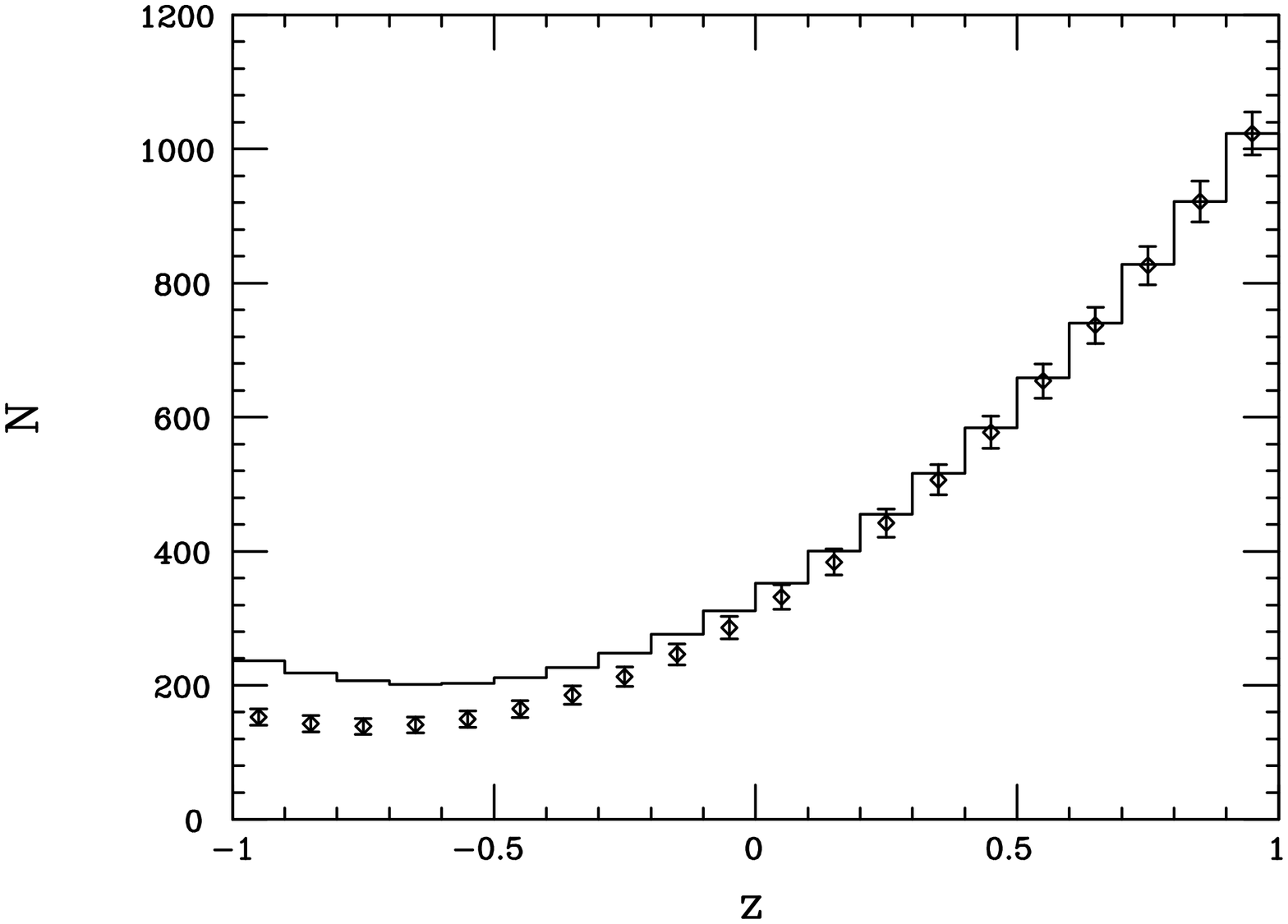,height=5.8cm,width=6.6cm,angle=-90}
\hspace*{-5mm}
\psfig{figure=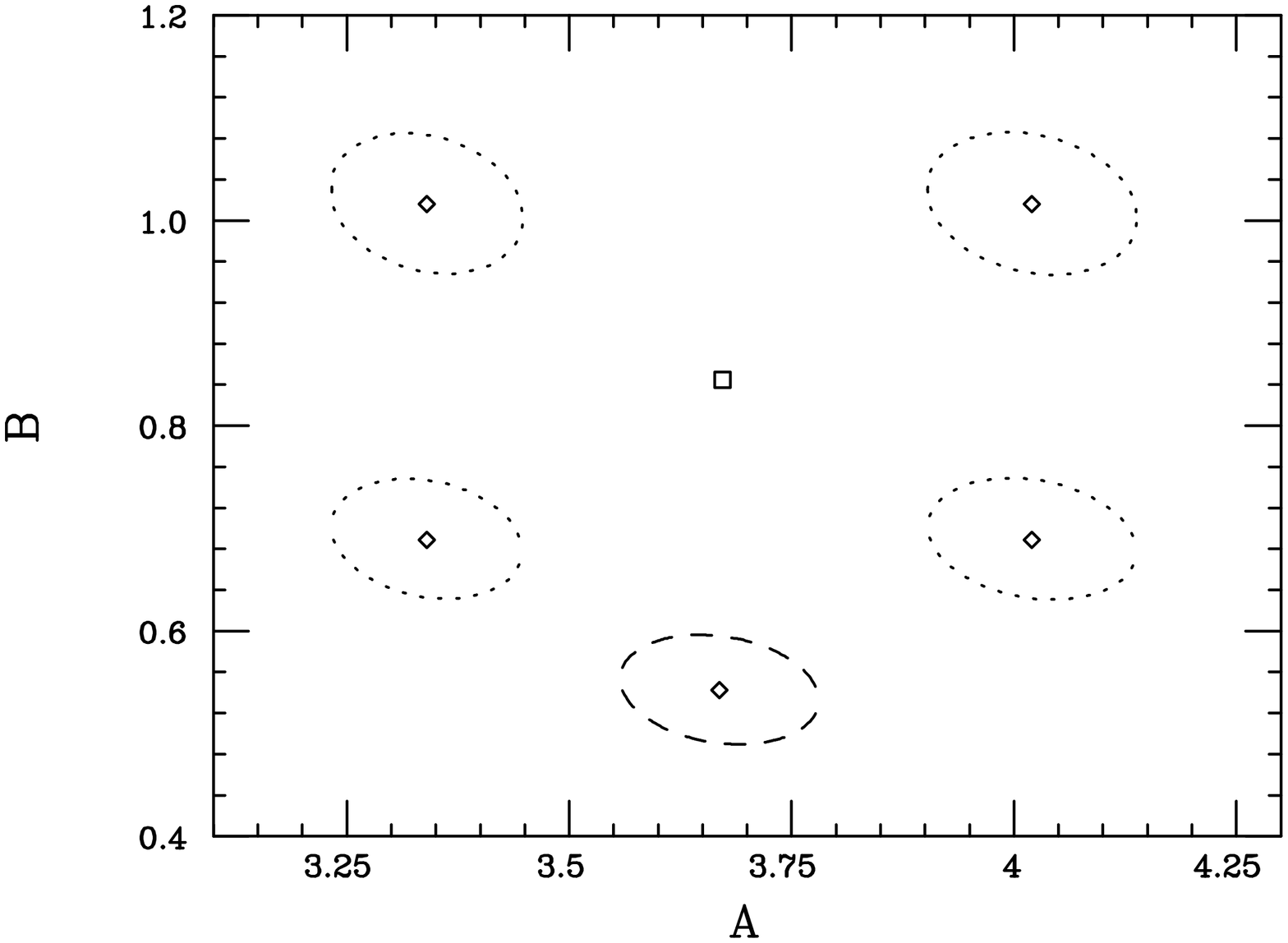,height=5.8cm,width=6.6cm,angle=-90}}
\vspace*{-1cm}
\caption{(Left) Binned angular distribution for the process 
$e^+e^-\to \mu^+\mu^-$ or 
$\tau^+\tau^-$ at a 1 TeV NLC in the SM (histogram) and for the 
case where a 3 TeV $\tilde \nu$ with $\lambda=0.5$ exchanged in 
the $t$-channel also contributes assuming an integrated luminosity of 
$L=150~fb^{-1}$. (Right) $95\%$ CL fits to the values of $A$ and $B$ for the 
data generated 
with $\tilde \nu$ exchange(dashed region) and for the data generated for the 
four typical choices of $Z'$ couplings(dots) employed in Fig. 2. The 
SM result is represented 
by the square in the center of the figure while the diamonds are the locations 
of the best fits.}
\end{figure}
\vspace*{0.4mm}
\vspace*{-0.5cm}
\nn
\begin{figure}[htbp]
\centerline{
\psfig{figure=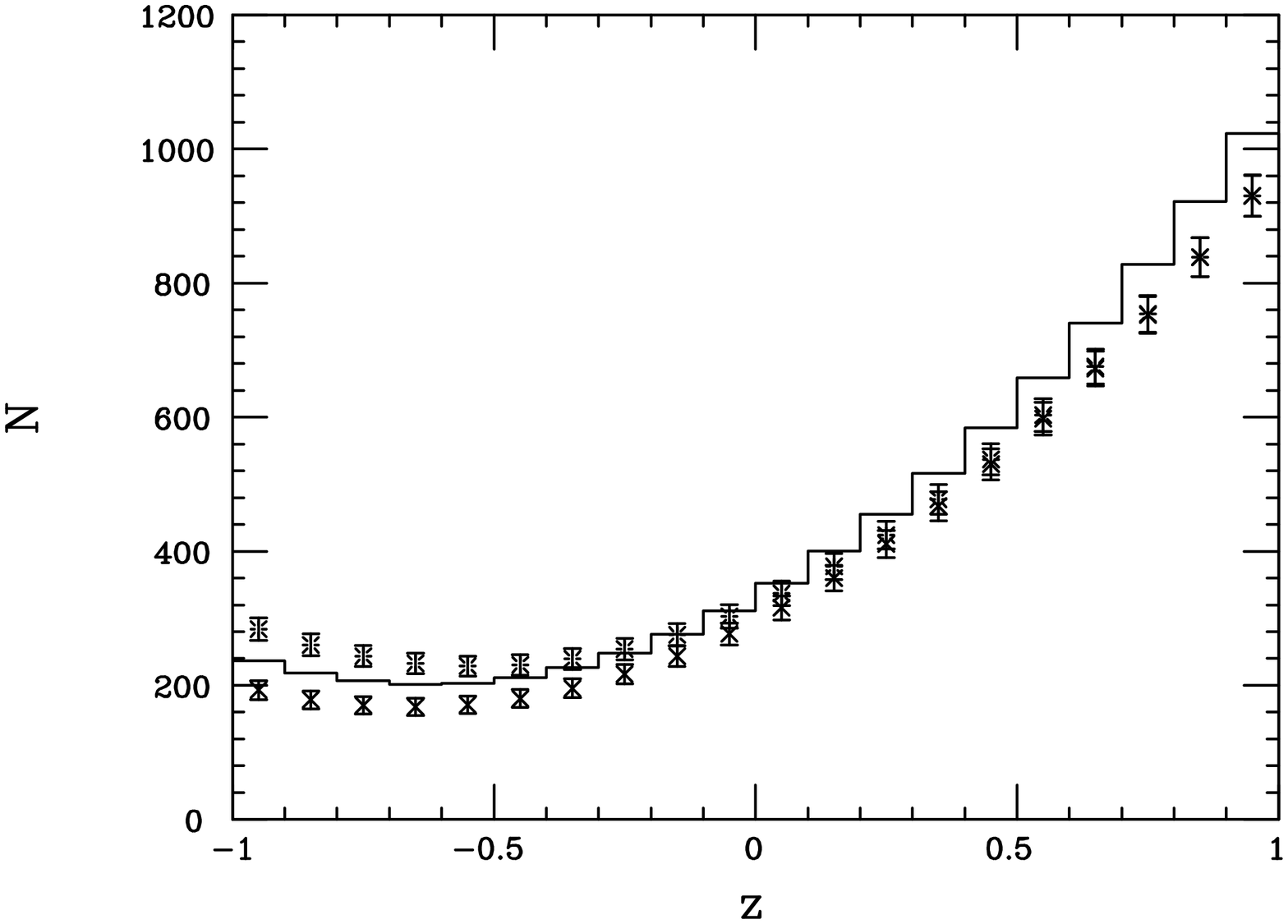,height=5.8cm,width=6.6cm,angle=-90}
\hspace*{-5mm}
\psfig{figure=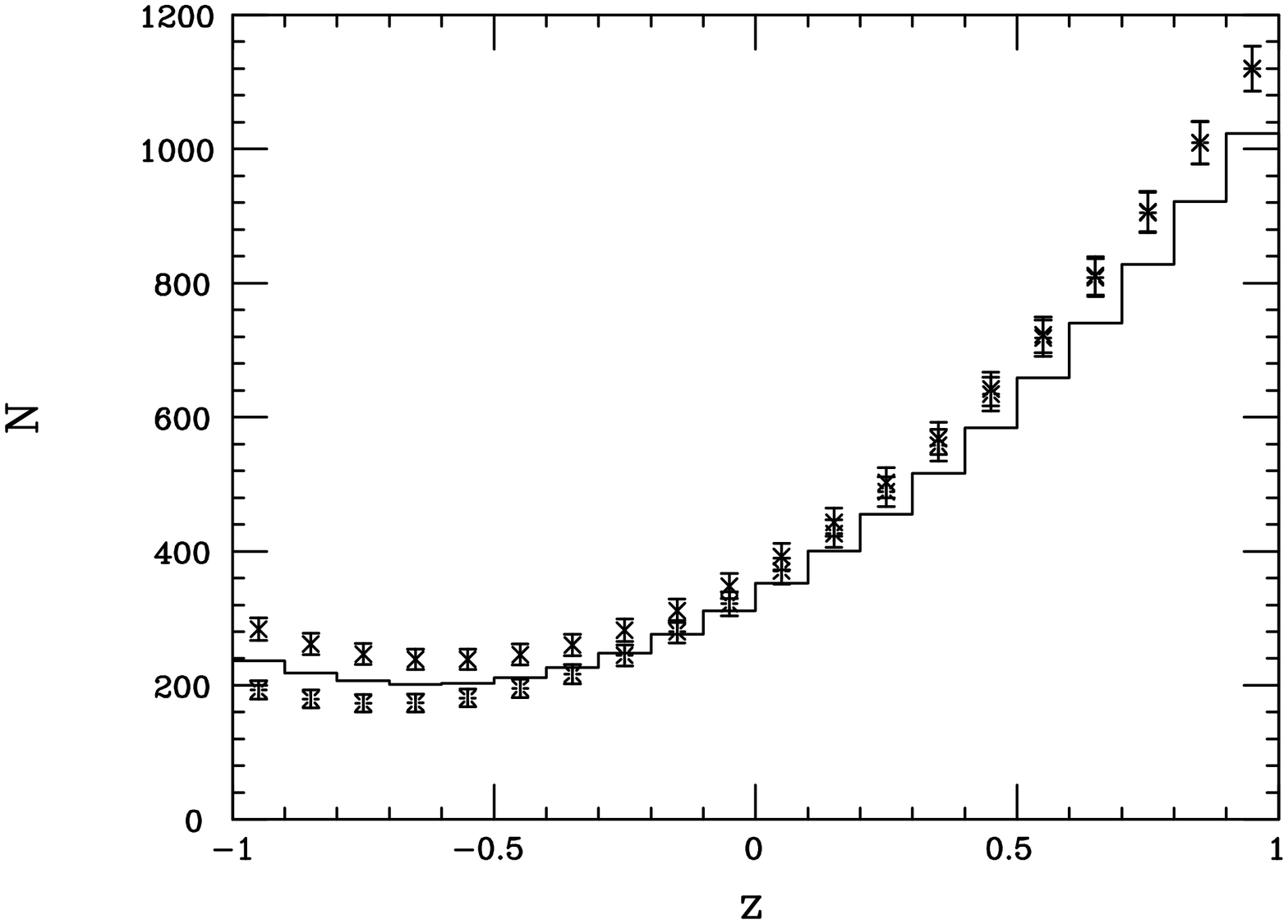,height=5.8cm,width=6.6cm,angle=-90}}
\vspace*{-1cm}
\caption{Same as the left panel of the 
previous figure but now including a 3 TeV $Z'$ exchange 
in the $s$-channel. The magnitude of all $Z'$ couplings is taken to be be the 
same value, \ie, $|g^{e\prime,f\prime}_{L,R}|=0.3c$, for purposes of 
demonstration. In the left panel, the relative signs of ($g^{e\prime}_{L}, 
~g^{e\prime}_{R},~g^{f\prime}_{L},~g^{f\prime}_{R}$ are chosen to be 
$(+,-,+,-)[(+,+,+,+)]$ for the upper[lower] series of data points, while in 
the right panel they correspond to the choices $(+,-,-,+)[(+,+,-,-)]$ for 
the upper[lower] series, respectively.}
\end{figure}
\vspace*{0.4mm}

In case ($ii$) where the $\tilde \nu$ is exchanged in the $s$-channel, the 
angular distribution gets modified to $\sim A(1+z)^2+B(1-z)^2+C$ with $A,B$ 
taking on their SM values and $C$ being a constant for fixed $s$. While $Z'$ 
exchange will lead to a good fit {\it for some} values of $A,B$, this will 
not happen in the $\tilde \nu$ case. The result of this analysis is shown in 
Fig. 3 where we see that the fit fails when $m_{\tilde \nu} \simeq 
(3.3-4.0)\lambda$ TeV. In the case where $\tau$ pairs are produced further 
sensitivity can be gained by constructing the spin-spin correlation, $B_{zz}$, 
as introduced by Bar-Shalom, Eilam and Soni{\cite {bumps}}. This quantity is 
unity in the SM as well as in extended gauge models but can be substantially 
smaller when 
$s$-channel scalars are present as shown in Fig. 3. Given the small 
statistical error anticipated at future colliders, $B_{zz}$ offers sensitivity 
to $\tilde \nu$ mass as large as $(4-6)\lambda$ TeV from the $\tau$ pair 
channel. In the case of the $\mu$ pair final state a similar asymmetry can 
be constructed provided both $e^\pm$ beams can be polarized:
\begin{equation}
A_{double}={\sigma(+,+)+\sigma(-,-)-\sigma(-,+)-\sigma(+,-)\over {\sigma(+,+)
+\sigma(-,-)+\sigma(-,+)+\sigma(+,-)}}\,. 
\end{equation}
where $\pm$ refer to the incoming $e^-$ and $e^+$ polarizations. $A_{double}$ 
takes on a fixed value in both the SM and in all $Z'$ models which is 
determined by the available beam polarizations. For example at a 1 TeV 
collider with $P_{e^-}=90\%$ and $P_{e^+}=65\%$, one finds $A_{double}=0.585$. 
However in the case of 
$\tilde \nu$ exchange the value of $A_{double}$ can be significantly reduced. 
The reach in this variable is found to be very comparable to that obtained 
from $B_{zz}$, \ie, $m_{\tilde \nu} \simeq (4-6)\lambda$ TeV.

\vspace*{-0.5cm}
\nn
\begin{figure}[htbp]
\centerline{
\psfig{figure=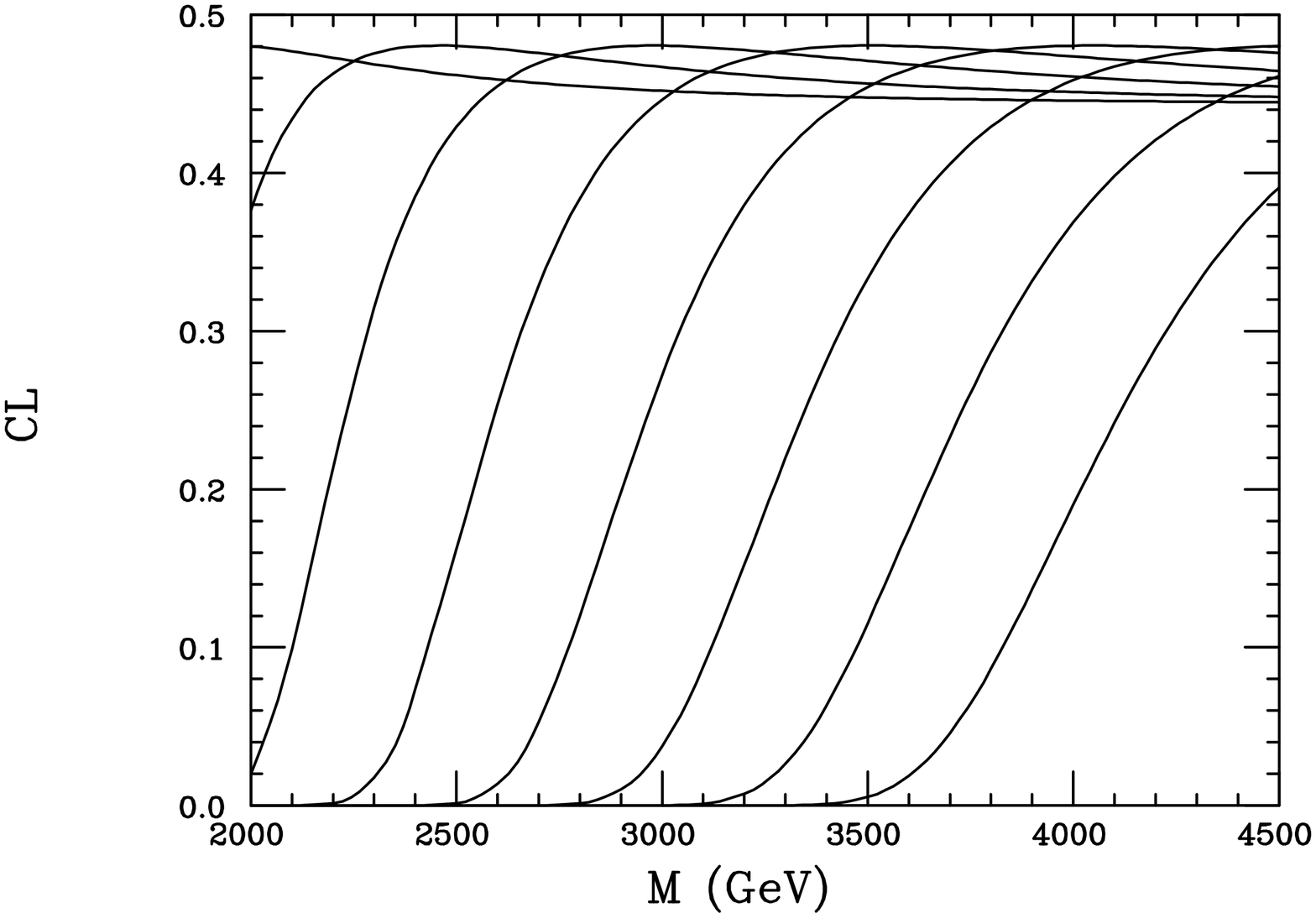,height=5.8cm,width=6.6cm,angle=-90}
\hspace*{-5mm}
\psfig{figure=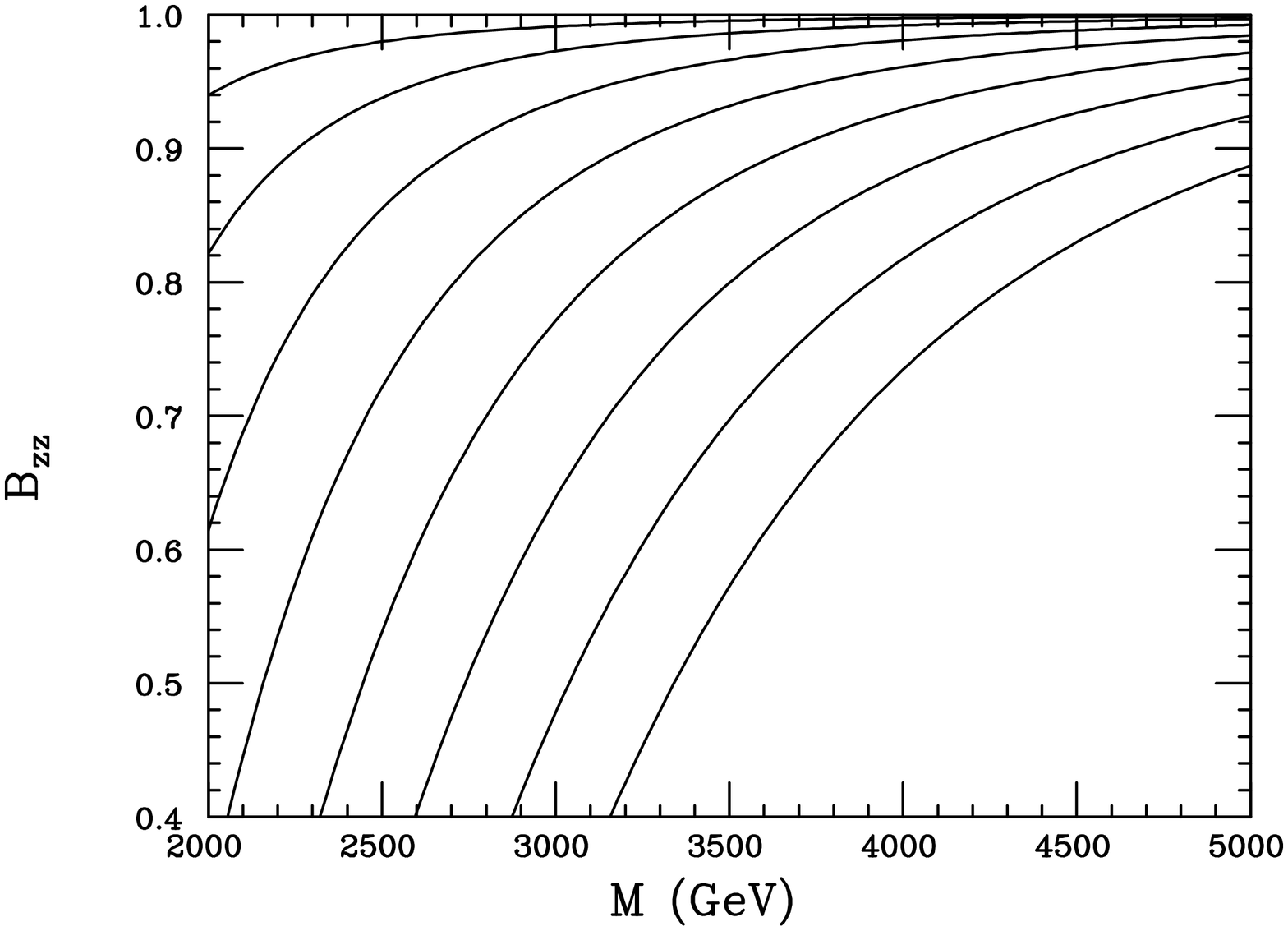,height=5.8cm,width=6.6cm,angle=-90}}
\vspace*{-1cm}
\caption{(Left) Average confidence level of the best fit to the parameters 
$A$ and $B$ as a function of the $\tilde \nu$ mass in the case of $s$-channel 
$\tilde \nu$ exchange for various values of the Yukawa 
coupling $\lambda$ in the range 0.3 to 1.0 in steps of 0.1 from top left to 
lower right. (Right) Double $\tau$ spin asymmetry at a 1 TeV NLC as a 
function of the 
$\tilde \nu$ mass for different values of the Yukawa coupling $\lambda$. 
From left to right, $\lambda$ varies from 0.3 to 1.0 in steps of 0.1 as in 
the left panel. In the case of either the SM or a $Z'$, $B_{zz}=1$.}
\end{figure}
\vspace*{0.4mm}

The final and most difficult case  to examine is ($iii$) Bhabha scattering 
since $s$- and $t$-channels exchanges are simultaneously present for both 
the $Z'$ and $\tilde \nu$ cases. To examine this cross 
section in any detail angular cuts are necessary to remove the photon pole; a 
cut $|z|<0.95$ will be assumed here which also removes a large SM background. 
Fig. 4 shows the contributions to Bhabha scattering from either a $Z'$ or 
$\tilde \nu$ exchange assuming an integrated luminosity of 150 $fb^{-1}$ at a 
1 TeV collider. Note that the $\tilde \nu$ exchange leads to an increase in 
the cross section in the backwards direction while $Z'$ exchange may either 
increase or decrease the cross section there. The distribution in the forward 
direction is little influenced by either type of new physics. In Fig. 4  we see 
that if the product of the left- and right-handed couplings to the $Z'$ is 
$>0$ we will not be able to separate a $Z'$ from $\tilde \nu$ contribution. 
This result remains true even if the double polarization asymmetry, 
$A_{double}$, is employed and this ambiguity remains. 

\vspace*{-0.5cm}
\nn
\begin{figure}[htbp]
\centerline{
\psfig{figure=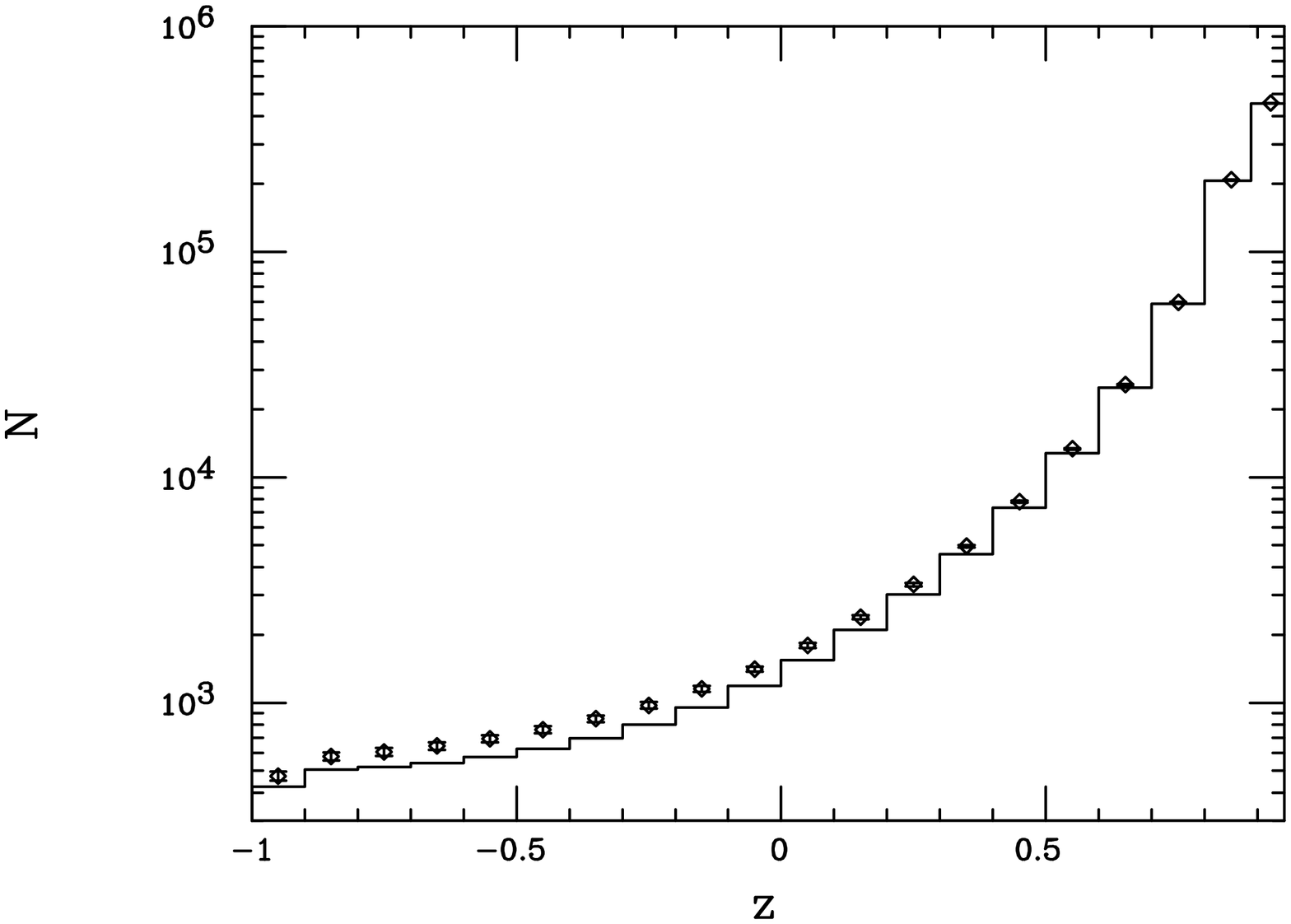,height=5.8cm,width=6.6cm,angle=-90}
\hspace*{-5mm}
\psfig{figure=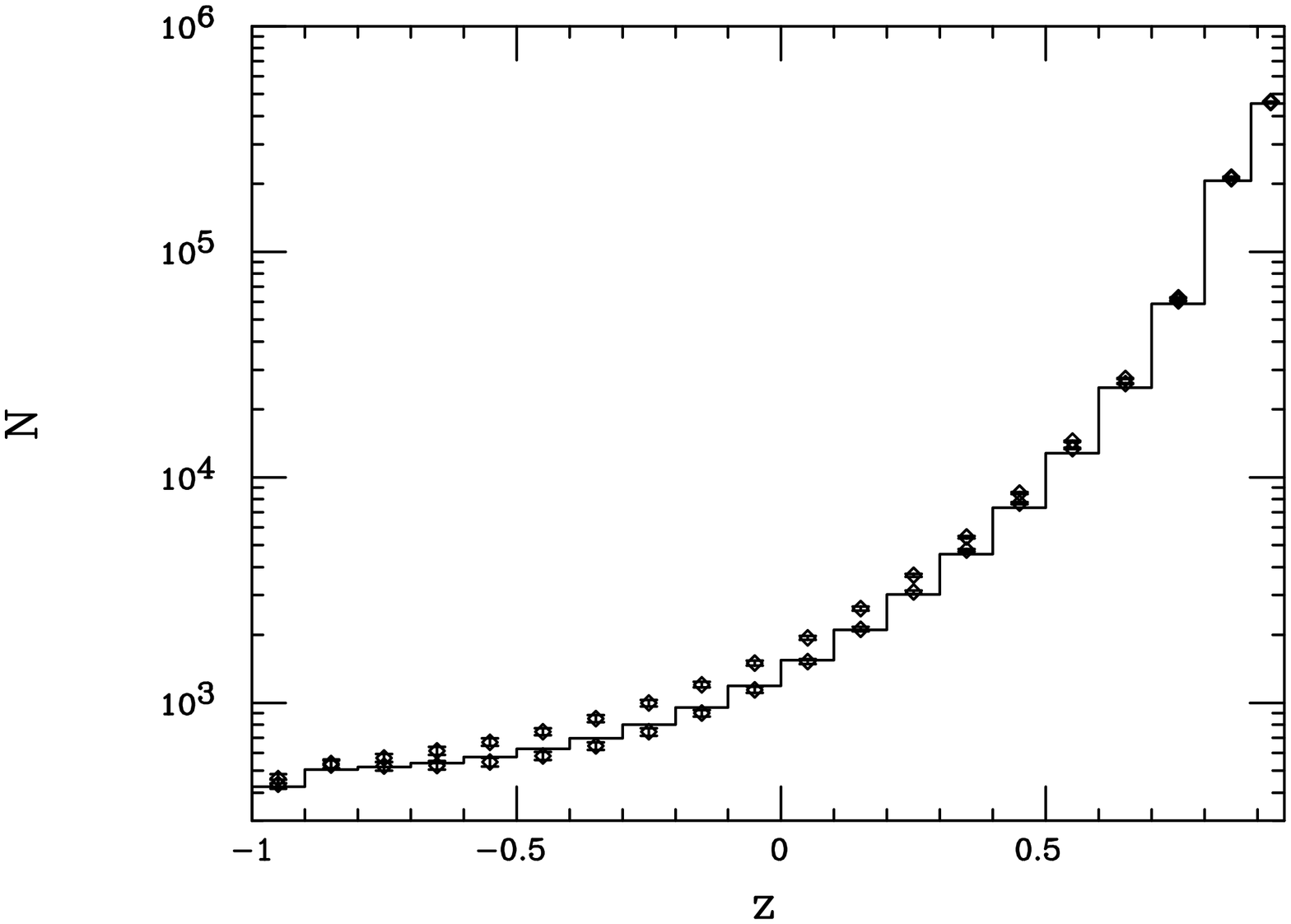,height=5.8cm,width=6.6cm,angle=-90}}
\vspace*{-1cm}
\caption{(Left) Same as left panel in Fig.1 but now for the case of Bhabha 
scattering. Angular cuts 
as described in the text have been employed to render the cross section 
finite in the forward direction.
(Right) Same as the left panel but now for a 3 TeV $Z'$ in comparison 
to the SM. The upper(lower) set of data points corresponds to 
$g^{e\prime}_L=g^{e\prime}_R=0.5c$($g^{e\prime}_L=-g^{e\prime}_R=0.5c$).}
\end{figure}
\vspace*{0.4mm}

In this talk we have considered the problem of how to distinguish two 
potential new physics scenarios from each other below the threshold for direct 
production of new particles at the NLC: $R$-parity violation and a extension 
of the SM gauge group by an additional $U(1)$ factor. Both kinds of new 
physics can lead to qualitatively similar alterations in SM cross sections,  
angular distributions and various asymmetries but differ in detail. These 
detailed differences provide the key to the two major weapons that are useful 
in accomplishing our task: ($i$) the angular distribution of the final state 
fermion and ($ii$) an asymmetry formed by polarizing both beams in the initial 
state, $A_{double}$. The traditional asymmetry, $A_{LR}$, formed when only a 
single beam is polarized, was shown~{\cite {tgr}} not to be useful for 
the case of purely 
leptonic processes we considered, but will be useful in an extension of the 
analysis to hadronic final states. This same analysis employed above can be 
easily extended to other new physics scenarios which involve the exchange on 
new particles~{\cite {jlh}} as in the case of massive graviton exchange in 
theories with compactified dimensions.

\section*{Acknowledgments}

The author would like to thank J.L. Hewett, S. Godfrey, P. Kalyniak, J. Wells, 
S. Bar-Shalom and H. Dreiner for discussions related to this work. 

\section*{References}

%
\def\MPL #1 #2 #3 {Mod.~Phys.~Lett.~{\bf#1},\ #2 (#3)}
\def\NPB #1 #2 #3 {Nucl.~Phys.~{\bf#1},\ #2 (#3)}
\def\PLB #1 #2 #3 {Phys.~Lett.~{\bf#1},\ #2 (#3)}
\def\PR #1 #2 #3 {Phys.~Rep.~{\bf#1},\ #2 (#3)}
\def\PRD #1 #2 #3 {Phys.~Rev.~{\bf#1},\ #2 (#3)}
\def\PRL #1 #2 #3 {Phys.~Rev.~Lett.~{\bf#1},\ #2 (#3)}
\def\RMP #1 #2 #3 {Rev.~Mod.~Phys.~{\bf#1},\ #2 (#3)}
\def\ZP #1 #2 #3 {Z.~Phys.~{\bf#1},\ #2 (#3)}
\def\IJMP #1 #2 #3 {Int.~J.~Mod.~Phys.~{\bf#1},\ #2 (#3)}

\end{document}